# Agentic Explainability at Scale: Between Corporate Fears and XAI Needs

XAI at Scale

Yomna Elsayed, PhD1*

Credo AI, yomna@credo.ai

Cecily K Jones, PhD

Credo AI, cecily@credo.ai

As companies enter the race for agentic AI adoption, fears surface around agentic autonomy and its subsequent risks. These fears compound as companies scale their agentic AI adoption with low-code applications, without a comparable scaling in their governance processes and expertise resulting in a phenomenon known as "Agent Sprawl". While shadow AI tools can help with agentic discovery and identification, few observability tools offer insights into the agents' configuration and settings or the decision-making process during agent-to-agent communication and orchestration. This paper explores AI governance professionals' concerns in enterprise settings, while offering design-time and runtime explainability techniques as suggested by AI governance experts for addressing those fears. Finally, we provide a preliminary prototype of an Agentic AI Card that can help companies feel at ease deploying agents at scale.

**ACM Reference:**

## 1 INTRODUCTION

Agentic AI deployments are rapidly expanding beyond isolated chatbots to complex, multi-agent systems, and enterprises are encouraging employees to innovate with low-code agents to unlock productivity gains. This growing footprint marks a shift from static, reactive, landscapes towards autonomous complex environments where agents reason, coordinate with other agents, invoke tools, and act within organizational workflows [1]. In multi-agent systems, AI does not operate in a vacuum but collaborates and communicates to achieve a goal. This shifts XAI needs from interpretability (understanding models) to auditability (tracking action).

Conversely, enterprise AI governance professionals are experiencing mounting anxiety over unchecked agent deployments, where agents consume excessive resources, duplicate existing functionality and operate without adequate monitoring. Thousands of unmonitored assets create opaque operational dependencies, resulting in "Agent Sprawl." This operational autonomy introduces profound challenges around accountability, predictability, and risk amplification where low-clearance agents may inadvertently access sensitive data through opaque chains of high-clearance dependencies [2,3]. While existing exploration has started to address foundational approaches to general explainability, less attention has been paid to XAI in enterprise environments where Agentic systems are deployed at scale, often by non-expert users. Prior work on explainable agents and agentic governance has largely focused on conceptual frameworks, architectural practices, or policy considerations, with limited evidence regarding how explainability is operationalized in large enterprise environments [4,2]. This paper focuses on Agentic XAI and how it can support governance, oversight, and trust in scaled, multi-agent enterprise environments.

## 2 RELATED WORK

Prior work in human-centered explainable AI (HCXAI) has argued that explainability extends beyond algorithmic transparency, and that who needs explanations and why, is as critical as technical mechanisms [5]. However, this discourse has largely centered on static AI systems leaving a gap in how human-centered explainability applies to autonomous, multi-agent, environments operating at enterprise scale. Unlike traditional systems that function as passive information processors, agentic systems introduce complexity in independent execution, adaptability, and the pursuit of complex, multi-step goals with limited direct supervision [2]. Although the benefits of embedded enterprise agentic workflows are clear, current implementation often fails to connect experimentation to governance. This creates a critical gap in how Agentic systems are controlled at scale [6].

In scaling environments, recent AI governance frameworks highlight that risk is no longer isolated to single model outputs. Instead, risk emerges through "cascading effects", in which hallucination and error are passed downstream, rapidly amplifying their impact [7]. These impacts are intensified through agent-to-agent interactions where feedback loops can reinforce errors over time, allowing risk to escalate beyond the speed at which human oversight and regulatory mechanisms can effectively intervene [3].

In the interconnected environment of organizational workflows, agents can dynamically connect with third-party tools or other agents, creating opaque chains of dependency where they may unintentionally coordinate while optimizing for local goals [7]. This leaves room for risk as third-party agents may influence or manipulate the primary agent's operations, making detection and control increasingly difficult as the number of autonomous interactions grows [2]. In these high stakes environments, low-fidelity explanations become insufficient for navigating liability and trust, and post-hoc explanations fail to provide the traceability, accountability, and system level visibility needed to understand how decisions emerge across interconnected systems [8]. Taken together, the current body of knowledge highlights a growing disconnect between the rapid adoption of agentic AI systems and the maturity of organizational frameworks designed to govern them. While prior research articulates the technical, architectural, and policy challenges associated with agentic autonomy and cascading risk, it offers limited insight into how challenges are understood and managed by practitioners operating in real-world enterprise settings [2,3]. There remains a lack of understanding of how domain experts conceptualize explainability requirements as agentic systems scale, interact, and evolve within organizational workflows.

## 3 METHODOLOGY

Mixed methods combining n=10 interviews with AI governance practitioners and domain experts in enterprises actively scaling agentic AI, with quantitative data from a large-scale survey of n=370 governance executives.

## 4 RESULTS

As companies transform their workflows from static AI models to agentic AI, the risks change in both complexity and volume. Low code solutions have enabled some companies to spawn thousands of agents without comparable governance mechanisms, creating a digital "Wild West". Unlike the carefully curated static AI models, these tools are now in the hands of non-technical employees, resulting in an explosion of "unmonitored assets". This is both a security and operational problem, where unchecked agents create duplicative work, consume resources and erode trust. As outputs shift from generated text to autonomous decisions and actions, the definition of XAI must evolve. Our research shows a distinctive move away from traditional model interpretability (e.g. feature attributions) toward systemic auditability. Governance professionals are prioritizing lineage of connectivity, and permission inheritance guardrails--shifting the focus from how the model thinks to controlling how the agent acts.

### 4.1 Governance Fears in Scaling Agentic AI

*4.1.1 Daisy-Chain reactions in agent connectivity*

Agents are autonomous entities that dynamically connect with other agents to access data, orchestrate decisions or act. In this environment, boundaries of control blurs. Practitioners are particularly worried about agent-to-agent interactions and how to control chain reactions in agent communication, whereby a company agent can "interact with a third-party agent, however that third party agent may also have access via MCP to a fifth party or fourth party without authorization". Governance professionals worry about data leakage or "adversarial persuasion occurring deep in the chain," invisible to the primary user. These deep connections also make it difficult to predict the ripple effect of one agentic change, leading to a loss of causal understanding and amplification of harms.

*4.1.2 Blast radius of permission inheritance*

The risk of uncontrolled connectivity is amplified by permission inheritance where agents operate with their creators' highest levels of access without corresponding oversight. In enterprises, agents usually run under the "user context" of creators. If a user with high-level clearance (e.g. access to Protected Health Information PHI) creates an agent, that agent can grant access to a lower-clearance agent. In this scenario, agents act like "Trojan horses" bypassing standard access controls. A governance professional in healthcare, noted "if you have PHI access and I don't, and you spin up an agent and then you give me access to the agent, because you have access to the PHI now I get access to the PHI by using that agent." The high clearance agent can expose through its outputs unauthorized data to the low-clearance agent. This creates a governance gap and widens the blast radius security breaches– if a high clearance agent is compromised, all its creator's entitlements are subsequently compromised.

*4.1.3 Accountability of agent actions*

Given the potential for agents to inherit vast permissions, accountability is an instrumental concern for governance professionals. Unlike LLMs which need a human to act on output, agents can act autonomously, even on misinformation. The spectrum of potential consequences can range from unauthorized data exfiltration (IP Leakage to competitors) to operational failure (irreversible database deletion). In large-scale organizations, it is important to maintain a trace of liability and accountability to ensure problems have a clear owner, are quickly fixed, and do not propagate. One practitioner demanded "If it's going to change settings on machinery, there has to be a human validation". Regulators argue that responsibility should lie with the deployer of the agent (the entity that put the model in action), not the model vendor. However, in practice practitioners note how it can be difficult for those deployers to maintain human oversight over every agent and their decisions. Practitioners are terrified of agents "tasking themselves indefinitely" without a clear human owner to hit the "kill switch. This fear is compounded when a failing agent has "write access", which can result in harm that cannot be rolled back.

### 4.2 Explainability Needs in Scaling Agentic AI

In a survey of 370 executives involved in AI governance decision making, "human oversight for high impact decisions", emerged as a critical factor for trusting and verifying autonomous decisions made by AI agents (40% rated highly critical). "Automated safeguards that block agents if they violate established policies" and "transparent audit trails tracing agent logic and sources" showed the broadest demand with 80.2% and 76.1% of respondents rating these as critical or highly critical, respectively. This indicates that traceability and preventative measures are becoming a baseline for Agentic AI adoption and trust. Qualitative data reflects a similar pattern, with governance professionals prioritizing identity controls and runtime traces to improve trust in agentic outputs. Interviews further emphasized the importance of design-time preventative measures alongside runtime observability, including agent inventories, agent cards, and dependency graphs. Specifically, design-time explainability interventions are necessary to address upfront organizational concerns around "Agent Sprawl," and "Permission Inheritance", while Runtime explainability is required to curtail mounting concerns around unpredictable daisy-chain reactions and agentic accountability.

*4.2.1 Design-time XAI*

**Inventory of Agents**

There was a near-universal consensus among practitioners that a prerequisite to Agentic explainability is an agent registry/inventory that provides visibility into what agents exist, who owns them, where they are deployed, and whether they are active. As one practitioner noted, "You have to start with a catalog, because the rest is very complex". In addition to addressing "Agent Sprawl," an inventory can strategically function as a dashboard to show leadership such as CIOs and CTOs, the extent of both adoption and risk. As one practitioner noted, CIOs must understand the governance implications of agentic adoption, particularly as thousands of agents are created without clear accountability or traceability.

**Agent Cards**

To complement the broader inventory, participants identified a need for Agent Cards, akin to model cards, which would function as a nutrition label detailing basic information about the agent. While traditional model cards document static information such as model purpose, architecture and training data, agent cards must uniquely document dynamic, and action-oriented elements like 'interacting entities' (e.g., tools and Model Context Protocols), its 'autonomy level', and its 'evidence of least privilege'.

Agent Cards would serve as a centralized resource for standardized information such as owner/deployer, developer, risk assessment, interacting entities, inputs and outputs, direct abilities, expected behaviors, and autonomy level (e.g., task-level, semi autonomous, or fully autonomous) to assign the appropriate level of governance. To address blast radius, practitioners also emphasized the importance of including evidence of least privilege with every agent card to ensure agents can only access the necessary data or tools without inheriting broad human user entitlements. By requiring standardized disclosures for agents, users can identify the authoritative versions to trust, drawing the line between what is a legitimate business tool and what is a shadow AI experiment. While Agent Cards help executives, like CIOs, understand ownership and high-level risk, AI builders and security practitioners rely on the cards' granular data to conduct technical assessments and apply technical controls, limiting an agent's blast radius and unauthorized actions.

**Dependency Graphs**

To address black-box daisy chains, dependency graphs serve as visual maps of agent connections and dependencies. By mapping connections, practitioners can determine what tools and agents deployed agents interact with (e.g. Agent A->Agent B->External tool). They are thus able to assess the extent of blast radii should an agent fail or be compromised (agents/assets that will be directly impacted or downstream). One practitioner expressed the need to visualize "the direct abilities of an application, what is second degree, third degree, here are the agents that my application is connected to, [being] able to double click into agents and see that graphical relationship." This layered visibility enables more proactive risk assessment and governance of agent ecosystems.

*4.2.2 Runtime XAI*

**Deep Observability**

As agents' complex and dynamic interactions can lead to consequential decisions, AI governance experts note that decision traces are often more important than logs. While logs document the inputs and outputs, decision traces document "chain-of-thought" and tool invocations made during execution. These traces constitute a "gold-mine" for debugging because "they show the reasoning steps the agent took to reach a decision". These observability traces address the concern around daisy-chain reactions, where agents can interact with "3rd or 4th party agents without authorization". They thus counter the black-box effect resulting from unauthorized actions which occur deep in the chain, such as adversarial persuasion or data leakage.

**Contextual Traceability**

In addition to decision traces, AI governance professionals expressed a need for understanding the context under which a decision is made, such as user meta data, agent authentication method, and the set of active permissions during executions. One practitioner noted, "traceability into the data... and the context [the agent] was running under when it made that decision" are crucial for trusting Agentic outputs. Having this visibility into context logs addresses the "blast radius" fear, allowing practitioners to assess if the agent's available permissions were properly scoped to the task (Least Privilege), and that no unauthorized lateral movement was attempted in the session.

**Operational Monitoring**

In addition to observability (post-hoc forensics), practitioners expressed a need for real-time monitoring and guardrails, allowing them to pull the "kill switch" on agents if they drifted from their assigned goals. "Where's the kill switch in the chain?", exclaimed one of our practitioners. Having those guardrails in place, would allow human operators to intervene, thus addressing the concerns around accountability and runaway autonomy when agents have write-access or can overstep their allowed scope of responsibilities.

## 5 DESIGN PROTOTYPE

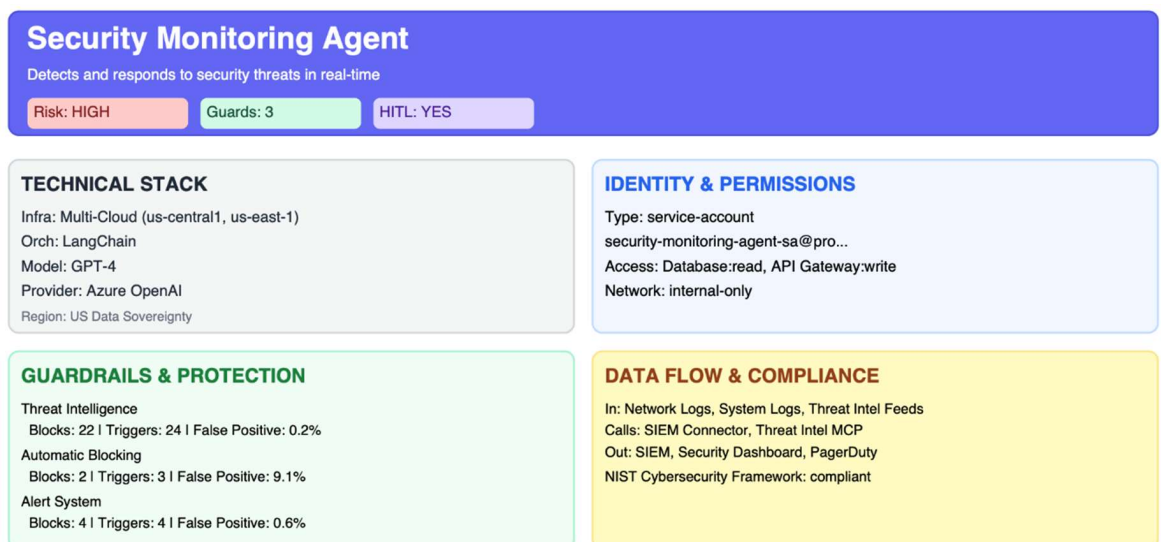

Figure 1: Agent Card for Security Monitoring Agent listing participants' explainability requirements for agentic trust

To provide stronger guarantees of completeness and accuracy, governance platforms should have the ability to automatically generate agent cards by pulling data directly from Agent platforms, and SDKs to extract architectural details and permissions. Additionally, by integrating with Agent evaluation suites, governance platforms can continually run live evaluations and push agent performance metrics onto the card to ensure the documentation remains highly accurate and dynamic rather than a static one-time document.

## 6 CONCLUSION

Governing agents at scale is a primary source of anxiety for AI governance professionals who are concerned about agent sprawl and unauthorized agent-to-agent interaction, resulting in uncontrolled daisy-chains, permission inheritance exploits and loss of accountability. To address this, companies need design-time explainability mechanisms such as centralized agent inventories, standardized agent cards and dependency graphs. Furthermore, they need run-time explainability that delivers deep observability, contextual traceability and operational monitoring, ensuring practitioners have oversight mechanisms needed to trust agentic output and safely scale AI.

## ACKNOWLEDGEMENTS

The authors would like to thank Credo AI and its AI governance experts for generously sharing their time, expertise, and insights. Their practical perspectives on agentic AI deployment and governance significantly informed the empirical and conceptual contributions of this work.